\documentstyle[twoside,fleqn,espcrc2]{article}
\input{psfig}

\def\VEV#1{\left\langle #1\right\rangle}
\def\sec{\ifmmode \,\, {\rm sec} \else sec \fi}
\def\eV {\ifmmode \,\, {\rm eV} \else eV \fi}
\def\keV{\ifmmode \,\, {\rm keV} \else keV \fi}
\def\MeV{\ifmmode \,\, {\rm MeV} \else MeV \fi}
\def\GeV{\ifmmode \,\, {\rm GeV} \else GeV \fi}
\def\TeV{\ifmmode \,\, {\rm TeV} \else TeV \fi}
\def\fm{\ifmmode \,\, {\rm fm} \else TeV \fi}
\def\pbarn{\ifmmode \,\, {\rm pb} \else pb \fi}
\def\km{\ifmmode {\rm km}\, \else km \fi}
\def\Mpc{\ifmmode {\rm Mpc}\, \else Mpc \fi}
\def\Gyr{\ifmmode {\rm Gyr}\, \else Gyr \fi}

\def\fun#1#2{\lower3.6pt\vbox{\baselineskip0pt\lineskip.9pt
  \ialign{$\mathsurround=0pt#1\hfil##\hfil$\crcr#2\crcr\sim\crcr}}}
\def\la{\mathrel{\mathpalette\fun <}}
\def\ga{\mathrel{\mathpalette\fun >}}

\def\sbar#1{\kern 0.8pt
        \overline{\kern -0.8pt #1 \kern -0.8pt}
        \kern 0.8pt}  % improved overbar

\def\meter{\ifmmode \,\, {\rm m} \else m \fi}
\def\yr {\ifmmode \,\, {\rm yr} \else yr \fi}

\def\sr{\ifmmode \,\, {\rm sr} \else sr \fi}

\def\hatn{{\bf \hat n}}
\newcommand{\AmS}{{\protect\the\textfont2
  A\kern-.1667em\lower.5ex\hbox{M}\kern-.125emS}}

\title{Cosmic Microwave Background Tests of Inflation}

\author{M. Kamionkowski\address{Columbia University, Department
        of Physics \\
	 538 West 120th Street, New York, NY
	10027~~USA}\thanks{Supported by D.O.E. contract DEFG02-92-ER
	40699, NASA NAG5-3091, and the Alfred P. Sloan
	Foundation.}}
       
\begin{document}

\begin{abstract}
Inflation provides a unified paradigm for understanding the
isotropy of the cosmic microwave background (CMB), the flatness
problem, and the origin of large-scale structure.  Although the
physics responsible for inflation is not yet well understood,
slow-roll inflation generically makes several predictions: a
flat Universe, primordial adiabatic density perturbations, and a
stochastic gravity-wave background. Inflation further predicts
specific relations between the amplitudes and shapes of the
spectrum of density perturbations and gravity waves.  There are
now excellent prospects for testing precisely these predictions
with forthcoming CMB temperature and polarization maps.  Here I
discuss these new CMB tests of inflation.
\end{abstract}

\maketitle

\section{INTRODUCTION}

Despite its major triumphs (the expansion, nucleosynthesis,
and the cosmic microwave background), the big-bang theory for
the origin of the Universe leaves several questions unanswered.
Chief amongst these is the horizon problem:  When cosmic
microwave background (CMB) photons last scattered, the age of
the Universe was roughly 100,000 years, much smaller than its
current age of roughly 10 billion years.  After taking into
account the expansion of the Universe, one finds that the angle
subtended by a causally connected region at the surface of last
scatter is roughly $1^\circ$. However, there are 40,000 square
degrees on the surface of the sky.  Therefore, when we look at
the CMB over the entire sky, we are looking at 40,000
disconnected regions of the Universe.  But quite remarkably,
each has the same temperature to roughly one part in $10^5$!

The most satisfying (only?) explanation for this is slow-roll
inflation \cite{inflation}, a
period of accelerated expansion in the early Universe driven by
the vacuum energy most likely associated with a symmetric phase
of a GUT Higgs field (or perhaps Planck-scale physics or
Peccei-Quinn symmetry breaking).  Although the physics
responsible for inflation is still not well understood,
inflation generically predicts (1) a flat Universe; (2) that
primordial adiabatic (i.e., curvature) perturbations are
responsible for the large-scale structure (LSS) in the Universe
today \cite{scalars}; and (3) a stochastic gravity-wave
background \cite{abbott}. More
precisely, inflation predicts a spectrum $P_s = A_s k^{n_s}$
(with $n_s$ near unity) of primordial density (scalar metric)
perturbations, and a stochastic gravity-wave background (tensor
metric perturbations) with spectrum $P_t =A_t \propto k^{n_t}$
(with $n_t$ small compared with unity).  Inflation further
uniquely predicts (4) specific relations between the
``inflationary observables,'' the amplitudes $A_s$ and $A_t$ and
spectral indices $n_s$ and $n_t$ of the scalar and tensor
perturbations \cite{steinhardt}.  The amplitude of the
gravity-wave background is
proportional to the height of the inflaton potential, and the
spectral indices depend on the shape of the inflaton potential.
Therefore, determination of these parameters would illuminate
the physics responsible for inflation.  

Until recently, none of these predictions could be tested with
precision.  Measured values for the density of the Universe span
almost an order of magnitude.  Furthermore, most do not probe
the possible contribution of a cosmological constant (or some
other diffuse matter component), so they do not address the
geometry of the Universe.  The only observable effects of a
stochastic gravity-wave background are in the CMB.  COBE
observations do in fact provide an upper limit to the tensor
amplitude, and therefore an inflaton-potential height near the
GUT scale.  However, there is no way to disentangle the scalar
and tensor contributions to the COBE anisotropy.

In recent years, it has become increasingly likely that
adiabatic perturbations are responsible for the
origin of structure.  Before COBE, there were numerous plausible
models for structure formation: e.g., isocurvature perturbations
both with and without cold dark matter, late-time or slow phase 
transitions, topological defects (cosmic strings or textures),
superconducting cosmic strings, explosive or seed models, a
``loitering'' Universe, etc.  However, after COBE, only
primordial adiabatic perturbations and topological defects were
still considered seriously.  And in the past few
months, some leading proponents of topological defects have
conceded that these models have difficulty accounting for the
origin of large-scale structure \cite{towel}.  

We are now entering an exciting new era, driven by new
theoretical ideas and developments in detector technology, in
which the predictions of inflation will be tested with
unprecedented precision.  It is even conceivable that early in
the next century, we will move from verification of inflation to
direct investigation of the high-energy physics responsible for
inflation.

The purpose of this talk is to review how forthcoming CMB
experiments will test several of these predictions.  I will
first review the predictions of inflation for density
perturbations and gravity waves.  I will then discuss how CMB
temperature anisotropies will test the inflationary predictions
of a flat Universe and a primordial spectrum of density
perturbations.  I will then review how a CMB polarization map
may be used to isolate the gravity waves and briefly review how
detection of these tensor modes may be used to learn about the
physics responsible for inflation.  I close with some brief
remarks about further testable consequences of inflation.

\section{INFLATIONARY OBSERVABLES}

Inflation occurs when the energy density of the Universe is
dominated by the vacuum energy $V(\phi)$ associated with some
scalar field $\phi$ (the ``inflaton'').  During this time, the
quantum fluctuations in $\phi$ produce classical scalar
perturbations, and quantum fluctuations in the spacetime metric
produce gravitational waves.  If the inflaton potential
$V(\phi)$ is given in units of $m_{\rm Pl}^4$, and the inflaton
$\phi$ is in units of $m_{\rm Pl}$, then the scalar and tensor
spectral indices are
\begin{eqnarray}
     1-n_s &=& { 1 \over 8\pi} \left( {V' \over V} \right)^2 -
     {1 \over 4 \pi} \left({V' \over V} \right)', \nonumber\\
     n_t &=& -{ 1 \over 8\pi} \left( {V' \over V} \right)^2. 
\label{spectralindices}
\end{eqnarray}
The amplitudes can be fixed by their contribution to $C_2^{\rm TT}$,
the quadrupole moment of the CMB temperature,
\begin{eqnarray}
     {\cal S} &\equiv & 6\, C_2^{{\rm TT},{\rm scalar}}= 33.2\,[V^3/(V')^2],
          \nonumber\\
     {\cal T} &\equiv & 6\, C_2^{{\rm TT},{\rm tensor}}= 9.2 \,V.
\label{amplitudes}
\end{eqnarray}
For the slow-roll conditions to be satisfied, we must have
\begin{eqnarray}
     (1 /16 \pi) (V'/V)^2 &\ll& 1, \\ 
     (1 /8\pi)(V''/V) & \ll & 1,
\label{slowrollconditions}
\end{eqnarray}
which guarantee that inflation lasts long enough to make the Universe
flat and to solve the horizon problem.

When combined with COBE results, current degree-scale--anisotropy and
large-scale-structure observations suggest that ${\cal T}/{\cal S}$ is less
than order unity in inflationary models, which restricts
$V\la 5\times 10^{-12}$.  If the consistency relation
${\cal T} / {\cal S} \simeq -7 n_t$ [implied by
Eqs. (\ref{spectralindices}) and (\ref{amplitudes})] holds, the
tensor spectrum must be nearly scale invariant ($n_t\simeq 0$).

\section{TEMPERATURE ANISOTROPIES}

The primary goal of CMB experiments that map the temperature as
a function of position on the sky is recovery of the
temperature autocorrelation function or angular power spectrum
of the CMB.  The fractional temperature perturbation
$\Delta T(\hatn)/T$ in a given direction $\hatn$ can be expanded
in spherical harmonics,
\begin{equation}
     {\Delta T(\hatn) \over T} = \sum_{lm} \, a_{(lm)}^{\rm T}\,
     Y_{(lm)}(\hatn),
\label{eq:Texpansion}
\end{equation}
where the multipole coefficients are given by
\begin{equation}
     a_{(lm)}^{\rm T} = \int\, d\hatn\, Y_{(lm)}^*(\hatn) \, {\Delta
     T(\hatn) \over T}.
\label{eq:alms}
\end{equation}
Statistical isotropy and homogeneity of the Universe imply that
these coefficients have expectation values $\VEV{ (a_{(lm)}^{\rm
T})^*
a_{(l'm')}^{\rm T}} = C_l^{\rm TT} \delta_{ll'} \delta_{mm'}$ when
averaged over the sky.  Roughly speaking, the multipole moments
$C_l^{\rm TT}$ measure the mean-square temperature difference
between two points separated by an angle $(\theta/1^\circ) \sim
200/l$.

\begin{figure*}[htbp]
\centerline{\psfig{file=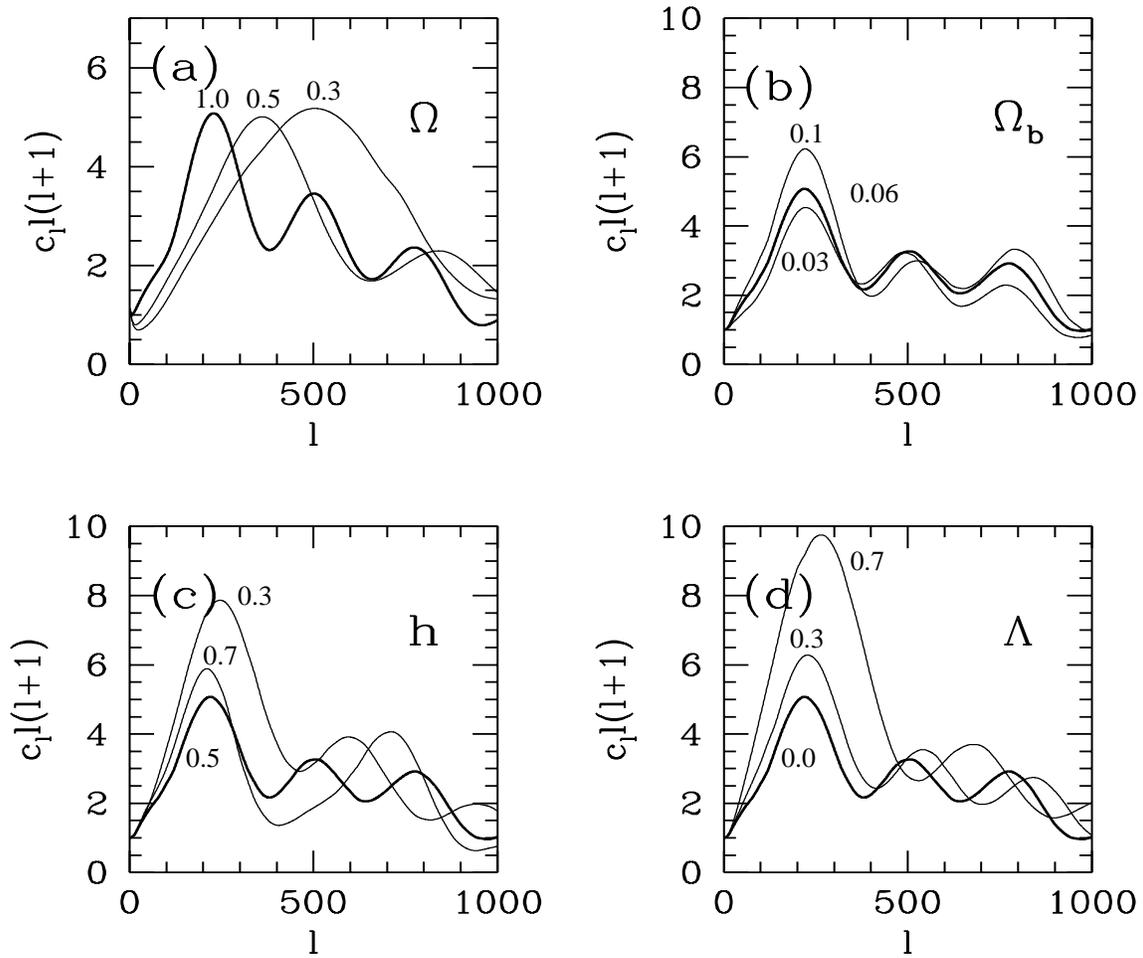,width=6in}}
\bigskip
\caption{
	  Theoretical predictions for CMB spectra as a function
	  of multipole moment $l$ for models with primordial
	  adiabatic perturbations.  In each case, the
	  heavy curve is that for the canonical standard-CDM values,
	  a total density $\Omega=1$, cosmological constant
	  $\Lambda=0$, baryon density $\Omega_b=0.06$, and
	  Hubble parameter $h=0.5$.  Each graph shows the effect
	  of variation of one of these parameters.  In (d),
	  $\Omega+\Lambda=1$.
}
\label{fig:models}
\end{figure*}

Predictions for the $C_l$'s can be made given a theory for
structure formation and the values of several cosmological
parameters.  Fig. \ref{fig:models} shows predictions for
models with primordial adiabatic perturbations.  The wriggles
come from oscillations in the photon-baryon fluid at the surface
of last scatter.  Each panel shows the effect of independent
variation of one of the cosmological parameters.  As
illustrated, the height, width, and spacing of the acoustic
peaks in the angular spectrum depend on  these (and other)
cosmological parameters.

In Ref. \cite{kamspergelsug}, we argued that these small-angle
CMB anisotropies can be used to determine the geometry of the
Universe.  The angle
subtended by the horizon at the surface of last scatter is
$\theta_H \sim \Omega^{1/2} \;1^\circ$, and the peaks in the CMB
spectrum are due to causal processes at the surface of last
scatter.  Therefore, the angles (or values of $l$) at which the
peaks occur determine the geometry of the Universe.  This is
illustrated in Fig. \ref{fig:models}(a) where the CMB spectra
for several values of $\Omega$ are shown.  As illustrated in the
other panels, the angular position of the first  peak is
relatively insensitive to the values of other undetermined (or
still imprecisely determined) cosmological parameters such as
the baryon density, the Hubble constant, and the cosmological
constant (as well as several others not shown such as the
spectral indices and amplitudes of the scalar and tensor spectra
and the ionization history of the Universe).  Therefore,
determination of the location of this first acoustic peak should
provide a robust measure of the geometry of the Universe.

The precision attainable is ultimately limited by
cosmic variance and practically by the finite angular resolution,
instrumental noise, and partial sky coverage in a realistic CMB
mapping experiment.  Taking these considerations into account,
my collaborators and I showed that future satellite missions may
potentially determine $\Omega$ to better than 10\% {\it after}
marginalizing over all other undetermined parameters (we
considered 7 more parameters in addition to the 4 shown in
Fig. \ref{fig:models}), and better than 1\% if the other parameters can be
fixed by independent observations or assumption \cite{jkksone}.
This would be far more accurate than any traditional
determinations of the geometry.

We also found that the CMB should provide determinations of
the cosmological constant and baryon density far more precise
than those from traditional observations \cite{jkkstwo}.  If there is more
nonrelativistic matter in the Universe than baryons can account
for---as suggested by current observations---it will become
increasingly clear with future CMB measurements.  Subsequent
analyses have confirmed these estimates with more precise
numerical calculations \cite{bet}.

Although these forecasts relied on the assumptions that
adiabatic perturbations were responsible for structure formation
and that reionization would not erase CMB anisotropies, these
assumptions have become increasingly
justifiable in the past few years.  As discussed above,
the leading alternative theories for structure formation now
appear to be in trouble, and recent detections of CMB
anisotropy at degree angular separations show that the effects
of reionization are small.  

NASA has recently approved the flight of a satellite mission,
the Microwave Anisotropy Probe (MAP) \cite{MAP} in the year 2000
to carry out these measurements, and ESA has approved the
flight of a subsequent more precise experiment, the Planck
Surveyor \cite{PLANCK}.  Therefore, it appears increasingly
likely that the inflationary prediction of a flat Universe will
be carried out precisely in the near future.

The predictions of a nearly scale-free spectrum of primordial
adiabatic perturbations will also be further tested with
measurements of small-angle CMB anisotropies.  The existence and
structure of the acoustic peaks shown in Fig. \ref{fig:models}
will provide an unmistakable signature of adiabatic
perturbations \cite{huwhite} and the spectral index $n_s$ can be determined
from fitting the theoretical curves to the data in the same way
that the density, cosmological constant, baryon density, and
Hubble constant are also fit \cite{jkkstwo}.

Temperature anisotropies produced by a stochastic gravity-wave
background would affect the shape of the angular CMB spectrum,
but there is no way to disentangle the scalar and tensor
contributions to the CMB anisotropy in a model-independent way.
Unless the tensor signal is large, the cosmic variance from the
dominant scalar modes will provide an irreducible limit to the
sensitivity of a temperature map to a tensor signal \cite{jkkstwo}.

\section{CMB POLARIZATION AND GRAVITY WAVES}

Although a CMB temperature map cannot unambiguously distinguish
between the density-perturbation and gravity-wave contributions
to the CMB, the two can be decomposed in a model-independent
fashion with a map of the CMB polarization
\cite{probe,ourpolarization,selzald}.  Suppose we
measure the linear-polarization ``vector'' $\vec P(\hatn)$ at
every point $\hatn$ on the sky.\footnote{Strictly
speaking, the linear polarization does not transform as a
vector, but the argument given here generalizes when one
describes the polarization state properly as a symmetric trace-free
$2\times2$ tensor.}  Such a vector field can be written as the
gradient of a scalar function $A$ plus the curl of a vector
field $\vec B$,
\begin{equation}
     \vec P(\hatn) \, = \, \vec \nabla A \, + \, \vec\nabla \times \vec
     B.
\label{eq:curl}
\end{equation}
The gradient (i.e., curl-free) and curl components can be
decomposed by taking the divergence or curl of $\vec
P(\hatn)$ respectively.  Density perturbations are scalar metric
perturbations, so they have no handedness.  They can therefore
produce no curl.  On the other hand, gravitational waves {\it
do} have a handedness so they can (and we have shown that they
do) produce a curl.  This therefore provides a way to detect the
inflationary stochastic gravity-wave background and thereby
test the relations between the inflationary observables.  It
should also allow one to determine (or at least constrain in the
case of a nondetection) the height of the inflaton potential.

More precisely, the Stokes parameters $Q(\hatn)$ and $U(\hatn)$
(where $Q$ and $U$ are measured with respect to the polar ${\bf
\hat\theta}$ and azimuthal ${\bf \hat \phi}$ axes) which specify
the linear polarization in direction $\hatn$ are components of a
$2\times2$ symmetric trace-free (STF) tensor, 
\begin{equation}
  {\cal P}_{ab}(\hatn)={1\over 2} \left( \begin{array}{cc}
   \vphantom{1\over 2}Q(\hatn) & -U(\hatn) \sin\theta \\
   -U(\hatn)\sin\theta & -Q(\hatn)\sin^2\theta \\
   \end{array} \right),
\label{whatPis}
\end{equation}
where the subscripts $ab$ are tensor indices.
Just as the temperature is expanded in terms of spherical
harmonics, the polarization tensor can be expanded,
\cite{ourpolarization}
\begin{eqnarray}
      {{\cal P}_{ab}(\hatn)\over T_0} &=&
      \sum_{lm} \Biggl[ a_{(lm)}^{{\rm G}}Y_{(lm)ab}^{{\rm
      G}}(\hatn) \nonumber \\
      & & +a_{(lm)}^{{\rm C}}Y_{(lm)ab}^{{\rm C}}(\hatn)
      \Biggr],
\label{Pexpansion}
\end{eqnarray}
in terms of the tensor spherical harmonics $Y_{(lm)ab}^{\rm G}$
and $Y_{(lm)ab}^{\rm C}$, which are a complete basis for the
``gradient'' (i.e., curl-free) and ``curl'' components of the
tensor field, respectively.  The mode amplitudes are given by
\begin{eqnarray}
a^{\rm G}_{(lm)}&=&{1\over T_0}\int d\hatn\,{\cal P}_{ab}(\hatn)\, 
                                         Y_{(lm)}^{{\rm G}
					 \,ab\, *}(\hatn),\cr 
a^{\rm C}_{(lm)}&=&{1\over T_0}\int d\hatn\,{\cal P}_{ab}(\hatn)\,
                                          Y_{(lm)}^{{\rm C} \,
					  ab\, *}(\hatn), 
\label{Amplitudes}
\end{eqnarray}
which can be derived from the orthonormality properties,
\begin{eqnarray}
\int d\hatn\,Y_{(lm)ab}^{{\rm G}\,*}(\hatn)
             Y_{(l'm')}^{{\rm
	     G}\,\,ab}(\hatn)&=&\delta_{ll'}\delta_{mm'},  \cr 
\int d\hatn\,Y_{(lm)ab}^{{\rm C}\,*}(\hatn)
             Y_{(l'm')}^{{\rm
	     C}\,\,ab}(\hatn)&=&\delta_{ll'}\delta_{mm'},  \cr 
\int d\hatn\,Y_{(lm)ab}^{{\rm G}\,*}(\hatn)
             Y_{(l'm')}^{{\rm C}\,\,ab}(\hatn)&=&0.
\label{Orthonormality}
\end{eqnarray}
Here $T_0$ is the cosmological mean CMB temperature and $Q$ and
$U$ are given in brightness temperature units rather than flux
units.   Scalar perturbations have no handedness.  Therefore,
they can produce no curl, so $a_{(lm)}^{\rm C}=0$ for scalar
modes.  On the other hand tensor modes {\it do} have a
handedness, so they produce a non-zero curl, $a_{(lm)}^{\rm C}
\neq0$.

A given inflationary model predicts that the $a_{(lm)}^{\rm X}$
are gaussian random variables with zero mean,
$\VEV{a_{(lm)}^{\rm X}}=0$  (for ${\rm X},{\rm X}' = \{{\rm
T,G,C}\}$) and covariance $\VEV{\left(a_{(l'm')}^{\rm X'}
\right)^* a_{(lm)}^{\rm X}} = C_l^{{\rm XX}'}
\delta_{ll'}\delta_{mm'}$.   Parity demands that
$C_l^{\rm TC}=C_l^{\rm GC}=0$.  Therefore the statistics of the
CMB temperature-polarization map are completely specified by the
four sets of moments, $C_l^{\rm TT}$, $C_l^{\rm TG}$, $C_l^{\rm
GG}$, and $C_l^{\rm CC}$.   Also, as stated above, only tensor modes
will produce nonzero $C_l^{\rm CC}$.  

\begin{figure*}[htbp]
\centerline{\psfig{file=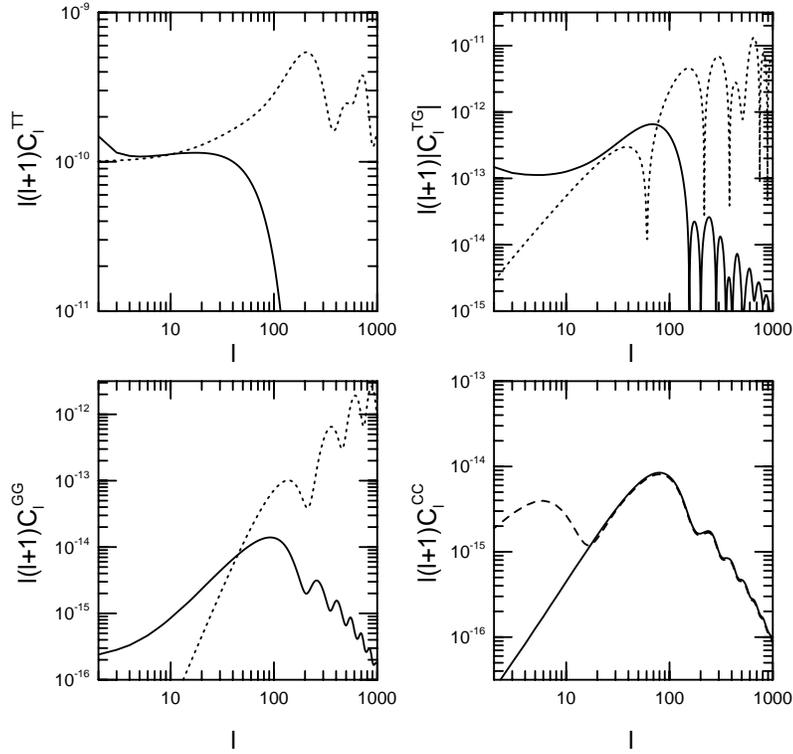,width=6in}}
\bigskip
\caption{
          Theoretical predictions for the four nonzero CMB
	  temperature-polarization spectra as a function
	  of multipole moment $l$.  The dotted curves are from
	  scalar perturbations in a COBE-normalized standard-CDM
	  model.  The solid curves are for a COBE-normalized
	  scale-invariant spectrum of tensor modes.  The dashed
	  curve in the CC panel shows the tensor spectrum for a
	  reionized model with optical depth $\tau=0.1$ to the
	  surface of last scatter.
       }
\label{clsplot}
\end{figure*}

To~illustrate, Fig.~\ref{clsplot} shows the four
temperature-polarization power spectra. The dotted curves
correspond to a COBE-normalized inflationary model
with cold dark matter and no cosmological constant
($\Lambda=0$), Hubble constant (in units of 100
km~sec$^{-1}$~Mpc$^{-1}$) $h=0.65$, baryon density
$\Omega_bh^2=0.024$, scalar spectral index $n_s=1$, no
reionization, and no gravitational waves.  The solid curves show
the spectra for a COBE-normalized stochastic gravity-wave
background with a flat scale-invariant spectrum ($h=0.65$,
$\Omega_b h^2=0.024$, and $\Lambda=0$) in a critical-density
Universe.   Note that the panel for $C_l^{\rm CC}$ contains no
dotted curve since scalar perturbations produce no C
polarization component.  The dashed curve in the CC panel shows
the tensor spectrum for a reionized model with optical depth
$\tau=0.1$ to the surface of last scatter.

As with a temperature map, the sensitivity of a polarization map
to gravity waves will be determined by the
instrumental noise and fraction of sky covered, and by the
angular resolution.  Suppose the detector sensitivity is $s$ and
the experiment lasts for $t_{\rm yr}$ years with an angular
resolution better than $1^\circ$.  Suppose further that we
consider only the CC component of the polarization in our
analysis.  Then the smallest tensor amplitude ${\cal T}_{\rm
min}$ to which the experiment will be sensitive at $1\sigma$ is
\cite{detectability}
\begin{equation}
     {{\cal T}_{\rm min}\over 6\, C_2^{\rm TT}}
      \simeq 5\times 10^{-4} \left( {s\over \mu{\rm K}\,\sqrt{\rm
      sec}} \right)^2 t_{\rm yr}^{-1}.
\label{CCresult}
\end{equation}
Thus, the curl component of a full-sky polarization map is
sensitive to inflaton potentials $V\ga 5 \times
10^{-15}t_{\rm yr}^{-1}$ $(s/\mu{\rm K}\, \sqrt{\rm sec})^2$.  
Improvement on current constraints with only the curl
polarization component requires a detector sensitivity
$s\la40\,t_{\rm yr}^{1/2}\,\mu$K$\sqrt{\rm sec}$.  For
comparison, the detector sensitivity of MAP will be $s={\cal
O}(100\,\mu$K$\sqrt{\rm sec})$.  However, Planck may conceivably
get sensitivities around $s=25\,\mu$K$\sqrt{\rm sec}$.

Even a small amount of reionization will significantly increase
the polarization signal at low $l$ \cite{reionization}, as shown
in the CC panel of
Fig.~\ref{clsplot} for $\tau=0.1$.  With such a level of
reionization, the sensitivity to the tensor amplitude is
increased by more than a factor of 5 over that in
Eq. (\ref{CCresult}).  This level of reionization (if not more)
is expected in cold dark matter models
\cite{kamspergelsug,blanchard,haiman}, so if anything,
Eq.~(\ref{CCresult}) provides a conservative estimate.

Furthermore, the estimate in Eq. (\ref{CCresult}) takes into
account only the information provided by the CC polarization
moments.  A complete likelihood analysis will fit the
temperature-polarization map to the four complete sets of
moments shown in Fig. \ref{clsplot}, and this will improve the
sensitivity significantly over that given in
Eq. (\ref{CCresult}) \cite{detectability}.

\section{DISCUSSION}

If MAP and Planck find a CMB temperature-anisotropy spectrum
consistent with a flat Universe and nearly--scale-free
primordial adiabatic perturbations, then the next step will be
to isolate the gravity waves with the polarization of the CMB.
If inflation has something to do with grand unification, then it
is possible that Planck's polarization sensitivity will be
sufficient to see the polarization signature of gravity waves.
However, it is also quite plausible that the height of the
inflaton potential may be low enough to elude detection by
Planck.  If so, then a subsequent experiment with better
sensitivity to polarization will need to be done.

Inflation also predicts that the distribution of primordial
density perturbations is gaussian, and this can be tested with
CMB temperature maps and with the study of the large-scale
distribution of galaxies.  Since big-bang nucleosynthesis
predicts that the baryon density is $\Omega_b \la 0.1$ and
inflation predicts $\Omega=1$, another prediction of inflation
is a significant component of nonbaryonic dark matter.  This can
be either in the form of vacuum energy (i.e., a cosmological
constant), and/or some new elementary particle.  Therefore,
discovery of particle dark matter could be interpreted as
evidence for inflation.

\end{document}